\documentclass[prl,aps,twocolumn,showpacs,nofootinbib,superscriptaddress]{revtex4-1}
\usepackage{epsfig}
\usepackage{amsmath, slashed}
\usepackage{color}

\newcommand{\uvec}{\boldsymbol}
\newcommand{\ud}{\mathrm{d}}

\allowdisplaybreaks
\begin{document}

\title{Charge Distributions of Moving Nucleons}

\author{C\'edric Lorc\'e}
\affiliation{CPHT, CNRS, Ecole Polytechnique, Institut Polytechnique de Paris, Route de Saclay, 91128 Palaiseau, France}
              
\begin{abstract}
Relativistic charge distributions of targets with arbitrary average momentum are introduced. They provide an interpolation between the usual Breit frame and infinite-momentum frame distributions. We find that Breit frame distributions can be interpreted from a phase-space perspective as internal charge quasi-densities in the rest frame of a localized target, without any relativistic correction. We show also that the apparent discrepancies between Breit frame and infinite-momentum frame distributions simply result from kinematical artifacts associated with spin. 
\end{abstract}

\pacs{14.20.Dh, 13.40.Gp}


\maketitle

Electromagnetic form factors (FFs) of nucleons and nuclei have been measured over the last decades to an impressive level of precision, see e.g.~\cite{Arrington:2006zm,Perdrisat:2006hj,Punjabi:2015bba,Pacetti:2015iqa}. They describe how the target reacts in an elastic scattering without getting excited, and contain therefore information about the internal distribution of charge and magnetization.

According to textbooks, FFs can be interpreted as Fourier transforms of charge and magnetization distributions. Since relativistic wave functions are frame dependent, Fourier transforms are often restricted to the Breit frame (BF)~\cite{Ernst:1960zza,Sachs:1962zzc}, where calculations formally yield the same results as in the non-relativistic domain. Concerns about the physical meaning of BF distributions have however been expressed~\cite{Yennie:1957,Breit:1964ga}, and their relation to genuine rest-frame distributions is usually thought to involve unclear and ambiguous relativistic corrections~\cite{Kelly:2002if}.

A strict density or probabilistic interpretation is tied to Galilean symmetry. In quantum field theory, it can only be justified when the momentum transfer remains small compared to the target inertia. Accordingly, the concept of rest-frame density is intrinsically limited by the Compton wavelength. One can however avoid these limitations in the infinite-momentum frame (IMF), where the target inertia becomes formally infinite~\cite{Fleming:1974af,Soper:1976jc,Burkardt:2000za, Miller:2010nz}. The price to pay is that the corresponding densities are now two-dimensional and appear to be distorted due to the motion of the target relative to the observer~\cite{Burkardt:2002hr,Carlson:2007xd}.

A phenomenological analysis of experimental data concluded that the center of the IMF charge distribution of the neutron is negative~\cite{Miller:2007uy}, in flagrant conflict with the rest-frame picture suggested by both gluon-exchange and meson-cloud models. Despite numerous efforts devoted to understanding this phenomenon, a fully convincing explanation has so far never been obtained.

Relaxing the requirement of a strict density interpretation, we show in the following that meaningful 2D charge distributions free of relativistic corrections can be defined for localized targets with arbitrary average momentum. They provide the natural interpolation between BF and IMF distributions and allow one to track down all distortions induced by the motion of the target. In particular, we find that a negative center in the neutron IMF distribution does not contradict the rest-frame picture and simply results from relativistic kinematical effects associated with spin.

We start with the observation that Lorentz symmetry implies that relativistic charge distributions are generally frame dependent. Their proper definition requires therefore to adopt a phase-space perspective. In a quantum theory, it has been known for a long time that the expectation value of any operator $\widehat O$ in a physical state $|\psi\rangle$ can nicely be expressed as~\cite{Wigner:1932eb,Hillery:1983ms}
\begin{equation}\label{PSampl}
\langle\widehat O\rangle_\psi=\int\frac{\ud^3P}{(2\pi)^3}\,\ud^3R\,\rho_\psi(\uvec R,\uvec P)\langle\widehat O\rangle_{\uvec R,\uvec P},
\end{equation}
where
\begin{equation}
\begin{aligned}
\rho_\psi(\uvec R,\uvec P)&\equiv\int\ud^3z\,e^{-i\uvec P\cdot\uvec z}\,\psi^*(\uvec R-\tfrac{\uvec z}{2})\psi(\uvec R+\tfrac{\uvec z}{2})\\
&=\int\frac{\ud^3q}{(2\pi)^3}\,e^{-i\uvec q\cdot\uvec R}\,\tilde\psi^*(\uvec P+\tfrac{\uvec q}{2})\tilde\psi(\uvec P-\tfrac{\uvec q}{2}).
\end{aligned}
\end{equation}
defines the quantum phase-space or Wigner distribution. Because of Heisenberg's uncertainty relations, Wigner distributions receive only a quasi-probabilistic interpretation: They give the quantum weight of finding the system at average position $\uvec R=\frac{1}{2}(\uvec x'+\uvec x)$ with average momentum $\uvec P=\frac{1}{2}(\uvec p'+\uvec p)$. Strict probabilistic interpretation is recovered under integration over position or momentum
\begin{align}
\int\ud^3R\,\rho_\psi(\uvec R,\uvec P)&=|\tilde\psi(\uvec P)|^2,\\
\int\frac{\ud^3P}{(2\pi)^3}\,\rho_\psi(\uvec R,\uvec P)&=|\psi(\uvec R)|^2.\label{Rprob}
\end{align}

Since wave-packet details have been factored out in Eq.~\eqref{PSampl},
\begin{equation}\label{intampl}
\langle\widehat O\rangle_{\uvec R,\uvec P}\equiv\int\frac{\ud^3\Delta}{(2\pi)^3}\,e^{i\uvec\Delta\cdot\uvec R}\langle\uvec P+\tfrac{\uvec\Delta}{2}|\widehat O|\uvec P-\tfrac{\uvec \Delta}{2}\rangle
\end{equation}
can be interpreted as the part associated with the internal structure of the system. When Galilean symmetry applies, $\langle\widehat O\rangle_{\uvec R,\uvec P}$ becomes $\uvec P$-independent and we recover a density interpretation owing to Eq.~\eqref{Rprob}. Although this formalism was originally developed in the non-relativistic context, it carries over to quantum field theory where position is understood in the Newton-Wigner sense~\cite{Newton:1949cq,Pavsic:2017orp}. 

Both initial and final states being on the mass shell $p'^2=p^2=M^2$, the four-momentum transfer $\Delta=p'-p$ is spacelike and orthogonal to the timelike average four-momentum $P=\frac{1}{2}(p'+ p)$. Its intrinsic meaning is then obtained in the class of elastic frames (EF) defined by the condition $\Delta^0=0$. In particular, the case $\uvec P=\uvec 0$ is known as the BF and corresponds from the phase-space perspective to the rest frame of the target localized around $\uvec R$. 

EF distributions were introduced in~\cite{Lorce:2017wkb,Lorce:2018egm} to study the frame dependence of the nucleon energy-momentum tensor. We define here in a similar way relativistic 2D distributions of the charge four-current (in units of the proton charge) as follows
\begin{equation}
\begin{aligned}
&J^\mu_\text{EF}(\uvec b_\perp;P_z)\equiv\int\ud r_z\,\langle\widehat j^\mu(r)\rangle_{\uvec R,P_z\uvec e_z}\\
&\quad=\int\frac{\ud^2\Delta_\perp}{(2\pi)^2}\,e^{-i\uvec\Delta_\perp\cdot\uvec b_\perp}\left[\frac{\langle p',s'|\widehat j^\mu(0)|p,s\rangle}{2P^0}\right]_{\Delta_z=0},
\end{aligned}
\end{equation}
where the $z$-axis is chosen for convenience along $\uvec P$ and $\uvec b_\perp=\uvec r_\perp-\uvec R_\perp$ are the impact parameter coordinates. The integration over the longitudinal coordinate in the first line ensures that the elastic condition $\Delta^0=0$ is satisfied when the target is moving $P_z\neq 0$. In the second line, we used translation invariance and we switched to momentum eigenstates  with covariant normalization $\langle p',s'|p,s\rangle=2p^0(2\pi)^3\delta^{(3)}(\uvec p'-\uvec p)\,\delta_{s's}$. The label $s$ ($s'$) denotes initial (final) canonical polarization.

Although it is true that relativistic densities cannot be defined in a model-independent way except in the IMF, we showed that unambiguous relativistic quasi-densities (or distributions) $J^\mu_\text{EF}(\uvec b_\perp;P_z)$ exist for any value of $P_z$. They are time independent (reflecting the fact that the target does not get excited during an elastic process in the first Born approximation) and they can be extended to 3D distributions in the BF. We note also that there are no limitations to the resolution since the average four-momentum is offshell $P^2=M^2-\Delta^2/4$, so that the constraint $|\uvec\Delta|<P^0$ is always satisfied.

For a spin-$0$ target, Lorentz symmetry implies that the generic off-forward matrix elements of the charge four-current operator can be written as
\begin{equation}
\langle p'|\widehat j^\mu(0)|p\rangle=2P^\mu F(Q^2)
\end{equation}
with $F(Q^2)$ a Lorentz-invariant function of $Q^2=-\Delta^2$. The corresponding relativistic 2D charge distribution takes the simple form
\begin{equation}
J^0_\text{EF}(\uvec b_\perp;P_z)=\int\frac{\ud^2\Delta_\perp}{(2\pi)^2}\,e^{-i\uvec\Delta_\perp\cdot\uvec b_\perp}F(\uvec\Delta^2_\perp)
\end{equation}
and appears to be independent of $P_z$. It is therefore the same in both the BF and the IMF, confirming that it is free of relativistic corrections. Since from a phase-space perspective the BF is the target rest frame, there is indeed no need to boost the system to set either $\uvec p$ or $\uvec p'$ to zero. Moreover, Lorentz contraction effects are automatically taken into account by the combination $|p\rangle/\sqrt{2p^0}$.

In the case of a spin-$\frac{1}{2}$ target like the nucleon, the generic off-forward matrix elements of the charge four-current operator
\begin{equation}\label{nucleoncurrent}
\langle p',s'|\widehat j^\mu(0)|p,s\rangle=\overline u(p',s')\Gamma^\mu(P,\Delta)u(p,s)
\end{equation}
are usually parametrized in terms of the Dirac and Pauli FFs
\begin{equation}
\Gamma^\mu(P,\Delta)=\gamma^\mu F_1(Q^2)+\frac{i\sigma^{\mu\nu}\Delta_\nu}{2M}\,F_2(Q^2).
\end{equation}
We find that these amplitudes can be interpreted in a more transparent way using the alternate but equivalent parametrization ($\epsilon_{0123}=+1$)
\begin{equation}\label{nucleoncurrentGEGM}
\begin{aligned}
&\Gamma^\mu(P,\Delta)=\\
&\quad\frac{MP^\mu }{P^2}\,G_E(Q^2)+\frac{i\epsilon^{\mu\alpha\beta\lambda}\Delta_\alpha P_\beta\gamma_\lambda\gamma_5}{2P^2}\,G_M(Q^2),
\end{aligned}
\end{equation}
where $G_{E,M}(Q^2)$ are known as the Sachs FFs~\cite{Ernst:1960zza,Sachs:1962zzc}
\begin{equation}\label{SachsFF}
\begin{aligned}
G_E(Q^2)&=F_1(Q^2)-\tau F_2(Q^2),\\
G_M(Q^2)&=F_1(Q^2)+F_2(Q^2)
\end{aligned}
\end{equation}
with $\tau=Q^2/4M^2$. We indeed recognize~\eqref{nucleoncurrentGEGM} as the covariant version in momentum space of the well-known decomposition of the charge current into convection and magnetization currents $\uvec J=\rho\,\uvec v+\uvec\nabla\times\uvec M$~\cite{Yennie:1957}.

Identifying the origin of the coordinates with the average position of the target, the relativistic 3D charge distribution in the BF is given by
\begin{equation}\label{SachsJ0}
J^0_B(\uvec r)=\int\frac{\ud^3\Delta}{(2\pi)^3}\,e^{-i\uvec\Delta\cdot\uvec r}\,\frac{M}{P^0}\,G_E(\uvec\Delta^2),
\end{equation}
and the 3D current distribution $\uvec J_{\!B}(\uvec r)=\uvec\nabla\times\uvec M_{\!B}(\uvec r)$ arises from the curl of the magnetization distribution
\begin{equation}\label{SachsJ0}
\uvec M_{\!B}(\uvec r)=\frac{\uvec\sigma}{2M}\int\frac{\ud^3\Delta}{(2\pi)^3}\,e^{-i\uvec\Delta\cdot\uvec r}\,\frac{M}{P^0}\,G_M(\uvec\Delta^2),
\end{equation}
where $\uvec\sigma$ are the Pauli matrices. They differ from the conventional Sachs distributions~\cite{Sachs:1962zzc} by a kinematical factor $M/P^0=(1+\tau)^{-1/2}$. It is the same factor that appears explicitly in the differential elastic cross section in the first Born approximation~\cite{Yennie:1957,Hand:1963zz}
\begin{equation}\label{crossGEGM}
\frac{\ud\sigma}{\ud\Omega}=\left(\frac{\ud\sigma}{\ud\Omega}\right)_\text{\!Mott}\left[G_E^2(Q^2)+\frac{\tau}{\epsilon}\,G^2_M(Q^2)\right]\frac{1}{1+\tau},
\end{equation}
where $\epsilon$ and $\left(\frac{\ud\sigma}{\ud\Omega}\right)_\text{\!Mott}$ are, respectively, the virtual photon polarization and the Mott cross section including recoil effects in the lab frame. In a non-relativistic expansion, it gives rise to the famous Darwin-Foldy term $\propto \frac{\uvec\Delta^2}{8M^2}$~\cite{Yennie:1957,Friar:1975pp} which is traditionally excluded from the definition of a charge distribution in both the atomic and nuclear physics literature. From a relativistic perspective, the kinematical factor is however essential to ensure that the total charge
\begin{equation}
\int\ud^3r\,\langle\widehat j^0\rangle_{\uvec R,\uvec P}(r)=\frac{\langle P,s|\widehat j^0(0)|P,s\rangle}{2P^0}=G_E(0)
\end{equation}
behaves as a Lorentz scalar quantity~\cite{Friar:1975pp}. Although several authors recommended its inclusion, a consistent and fully relativistic definition of the charge distribution was still missing. Sachs distributions have then been adopted by default in the literature, and the question of the kinematical factor disappeared in the limbo of relativistic uncertainties plaguing their physical interpretation. We demonstrated that adopting a phase-space perspective clarifies now the situation.

\begin{figure}[b]
\includegraphics[width=0.75\hsize]{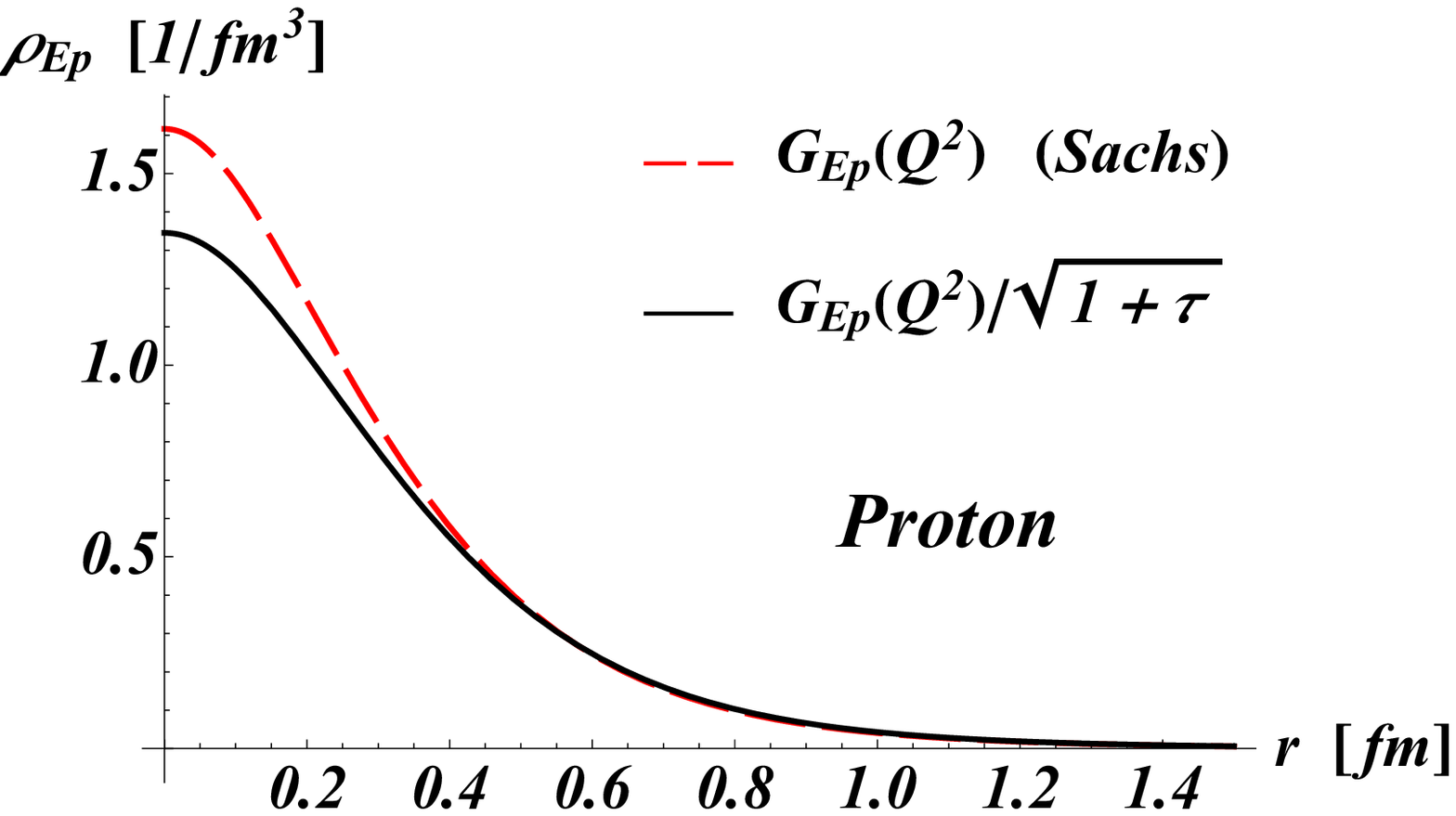}\vspace{.3cm}
\includegraphics[width=0.75\hsize]{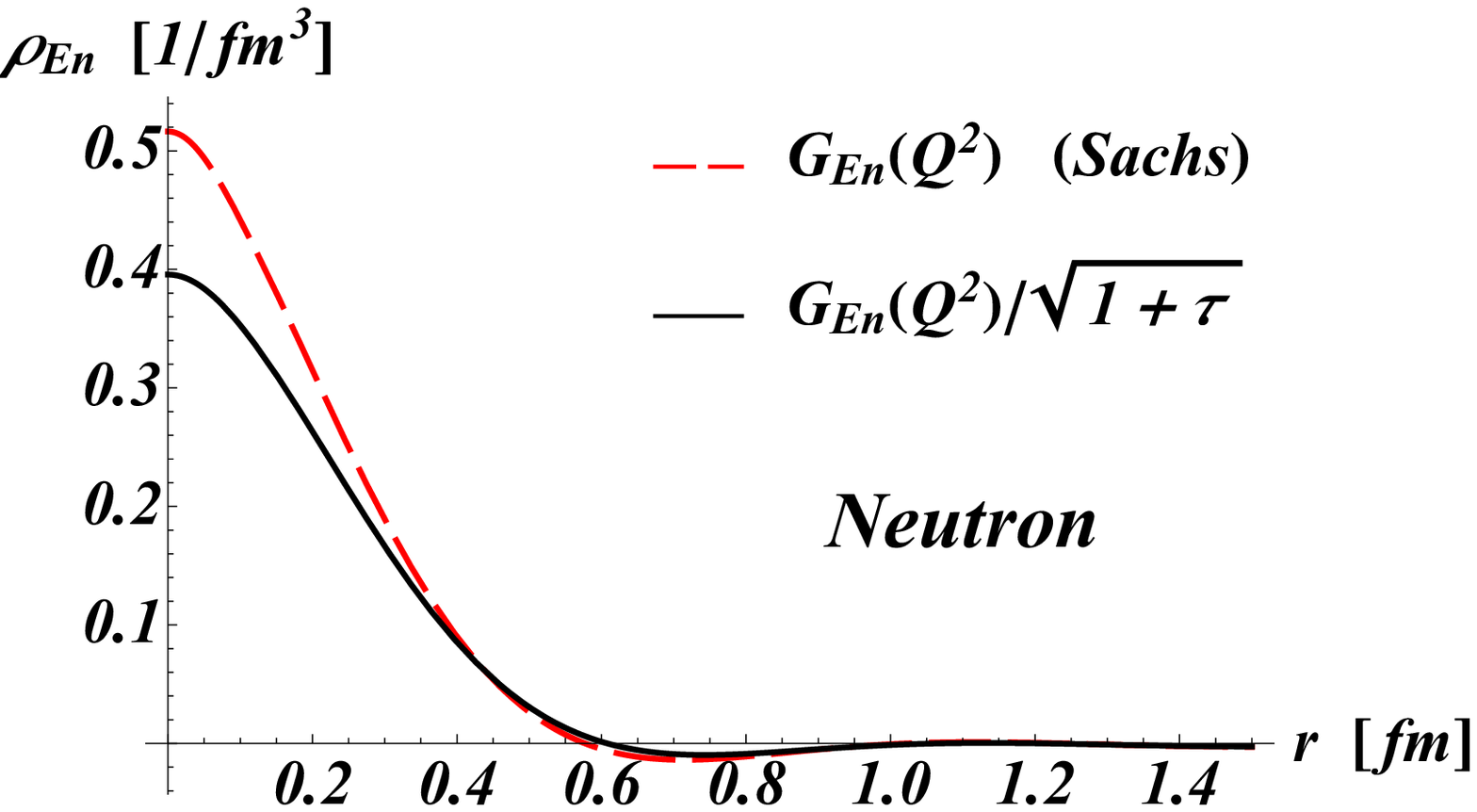}
\caption{(color online) Proton (top) and neutron (bottom) radial charge distributions in the Breit frame $\rho_E(r)=J^0_B(r\uvec e_r)$, excluding (dashed red) and including (solid black) the kinematical factor. Based on the parametrization from~\cite{Bradford:2006yz}.}\label{fig:SachsvsBF}
\end{figure}

In Fig.~\ref{fig:SachsvsBF} we compare the conventional Sachs charge distribution for the nucleons with the corresponding 3D quasi-density in the BF, using the phenomenological parametrization from~\cite{Bradford:2006yz}. We see that in both cases the center of the neutron charge distribution is positive, in agreement with the standard picture of a neutron fluctuating predominantly into a proton surrounded by a negatively charged pion cloud.

When the nucleon is moving, the charge distribution appears to be distorted due to relativistic kinematical effects associated with spin. We can indeed write in general~\cite{Jacob:1959at,Durand:1962zza}
\begin{equation}
\begin{aligned}
\langle p',s'|\widehat j^\mu(0)|p,s\rangle&=\sum_{s'_B,s_B}D^{*(j)}_{s'_Bs'}(p'_B,\Lambda)D^{(j)}_{s_Bs}(p_B,\Lambda)\\
&\qquad\times\Lambda^\mu_{\phantom{\mu}\nu}\langle p'_B,s'_B|\widehat j^\nu(0)|p_B,s_B\rangle,
\end{aligned}
\end{equation}
where $\langle p'_B,s'_B|\widehat j^\nu(0)|p_B,s_B\rangle$ is the BF amplitude, $\Lambda$ is the Lorentz boost from the BF to the generic frame, and $D^{(j)}$ is a Wigner rotation matrix for spin-$j$ targets. Setting $\mu=0$, we see that the EF charge distribution mixes both BF charge and current distributions. For a spinning target, the BF current does not vanish and the EF charge distribution receives a contribution from magnetization, see Eq.~\eqref{nucleoncurrentGEGM}. Note that this contribution does not change the total charge of the system, and hence just redistributes it in space. In particular, it induces a dipolar distortion of the charge distribution when the moving nucleon is transversely polarized~\cite{Burkardt:2002hr}. A similar phenomenon explains why the position of the center of inertia shifts sideways in a transversely polarized moving system~\cite{Moller:1949,Lorce:2018zpf}.

The second and more subtle effect comes from the Wigner rotation. It is a consequence of the non-commutativity of Lorentz boosts which makes polarization an observer-dependent concept. A given polarization in some frame appears rotated in another, explaining why simple relations among 3D parton distributions arise in spherically symmetric models~\cite{Lorce:2011zta}. Similarly, the BF charge (current) distribution is spin independent (dependent) in terms of the BF polarization, but appears to receive a spin-dependent (independent) contribution when described in terms of the polarization defined by an observer in another frame. 

Let us now focus on the unpolarized part of the EF charge distribution
\begin{equation}
\rho_E(b;P_z)\equiv\frac{1}{2}\,\textrm{Tr}[J^0_\text{EF}(\uvec b_\perp;P_z)],
\end{equation}
where the trace acts in polarization space. Since there is no preferred direction in the transverse plane, $\rho_E$ is axially symmetric and hence written as a function of the impact parameter $b=|\uvec b_\perp|$. Using explicit expressions for the Dirac bilinears~\cite{Lorce:2017isp}, we find for the convection and magnetization contributions
\begin{equation}
\rho^X_E(b;P_z)=\int_0^\infty\frac{\ud Q}{2\pi}\,Q\,J_0(Qb)\,\tilde\rho^X_E(Q;P_z),
\end{equation}
where $J_0$ is a cylindrical Bessel function and
\begin{equation}
\begin{aligned}
\tilde\rho^{conv}_E(Q;P_z)&=\frac{P^0+M(1+\tau)}{(P^0+M)(1+\tau)}\,G_E(Q^2),\\
\tilde\rho^{magn}_E(Q;P_z)&=\frac{\tau P^2_z}{P^0(P^0+M)(1+\tau)}\,G_M(Q^2)
\end{aligned}
\end{equation}
with $P^0=\sqrt{M^2(1+\tau)+P^2_z}$. In particular, we clearly see why the BF description $\tilde\rho_E(Q;0)=G_E(Q^2)/\sqrt{1+\tau}$ turns into the IMF description $\tilde\rho_E(Q;\infty)=F_1(Q^2)$: It is essentially due to a magnetization contribution arising from relativistic kinematical effects associated with spin. A similar analysis for the magnetization shows how $G_M(Q^2)/\sqrt{1+\tau}$ in the BF turns into $F_2(Q^2)$ in the IMF. This clarifies and completes the findings of~\cite{Rinehimer:2009yv}. Increasing the spin of the target will just increase the complexity of these effects~\cite{Carlson:2008zc,Alexandrou:2009hs,Lorce:2009bs}. 

\begin{figure}[b]
\includegraphics[width=0.51\hsize]{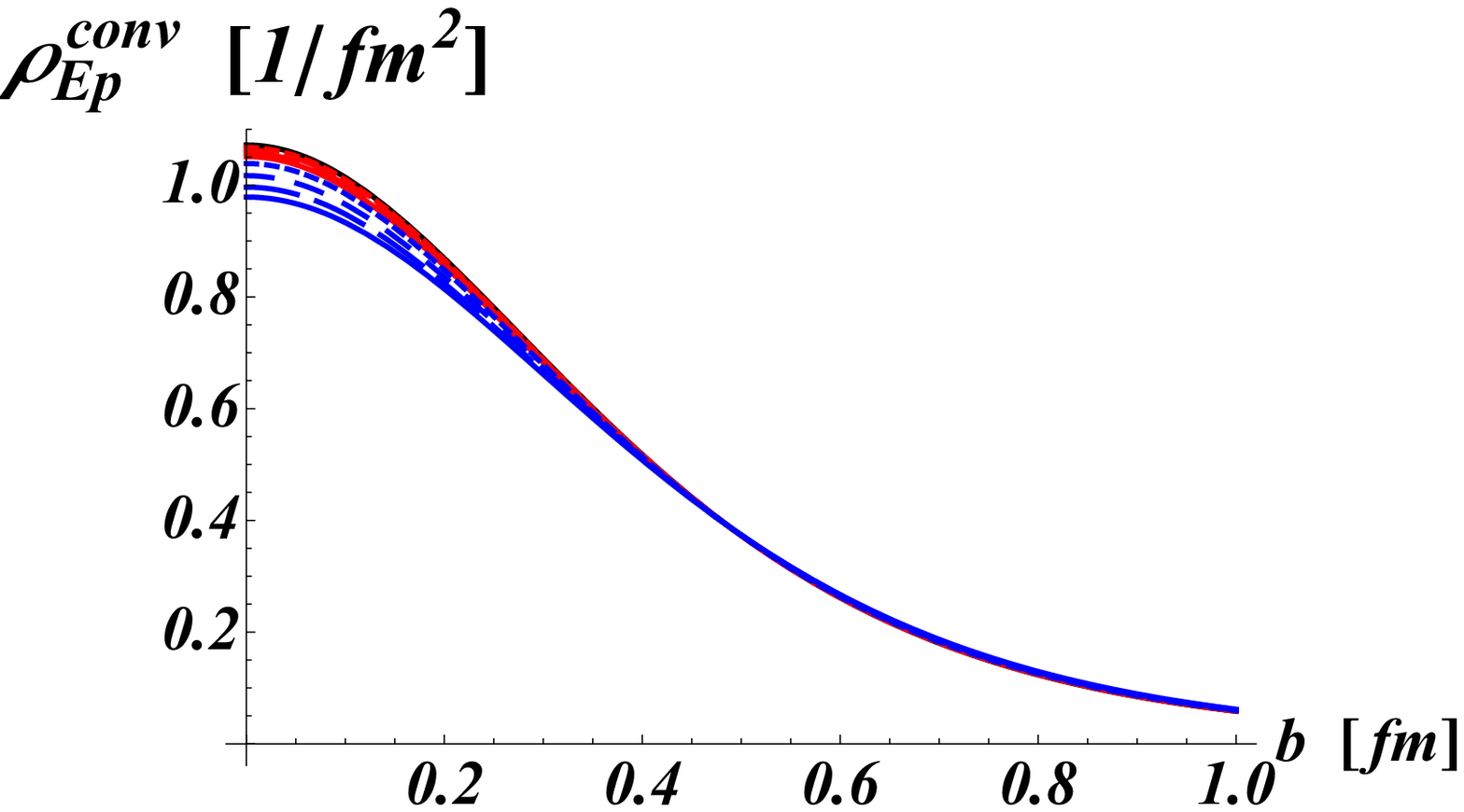}\hspace{-.4cm}
\includegraphics[width=0.51\hsize]{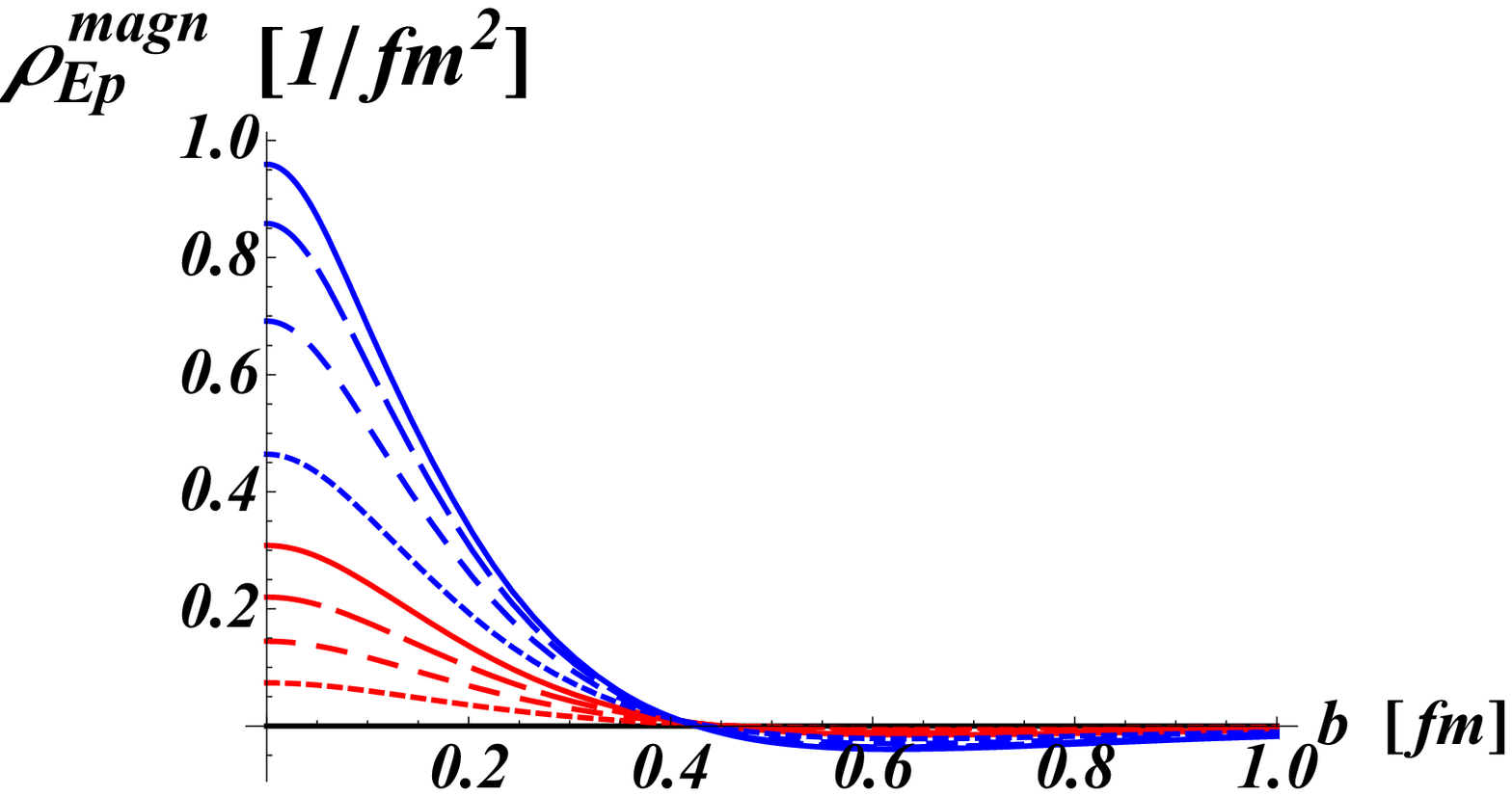}\vspace{.3cm}
\includegraphics[width=0.85\hsize]{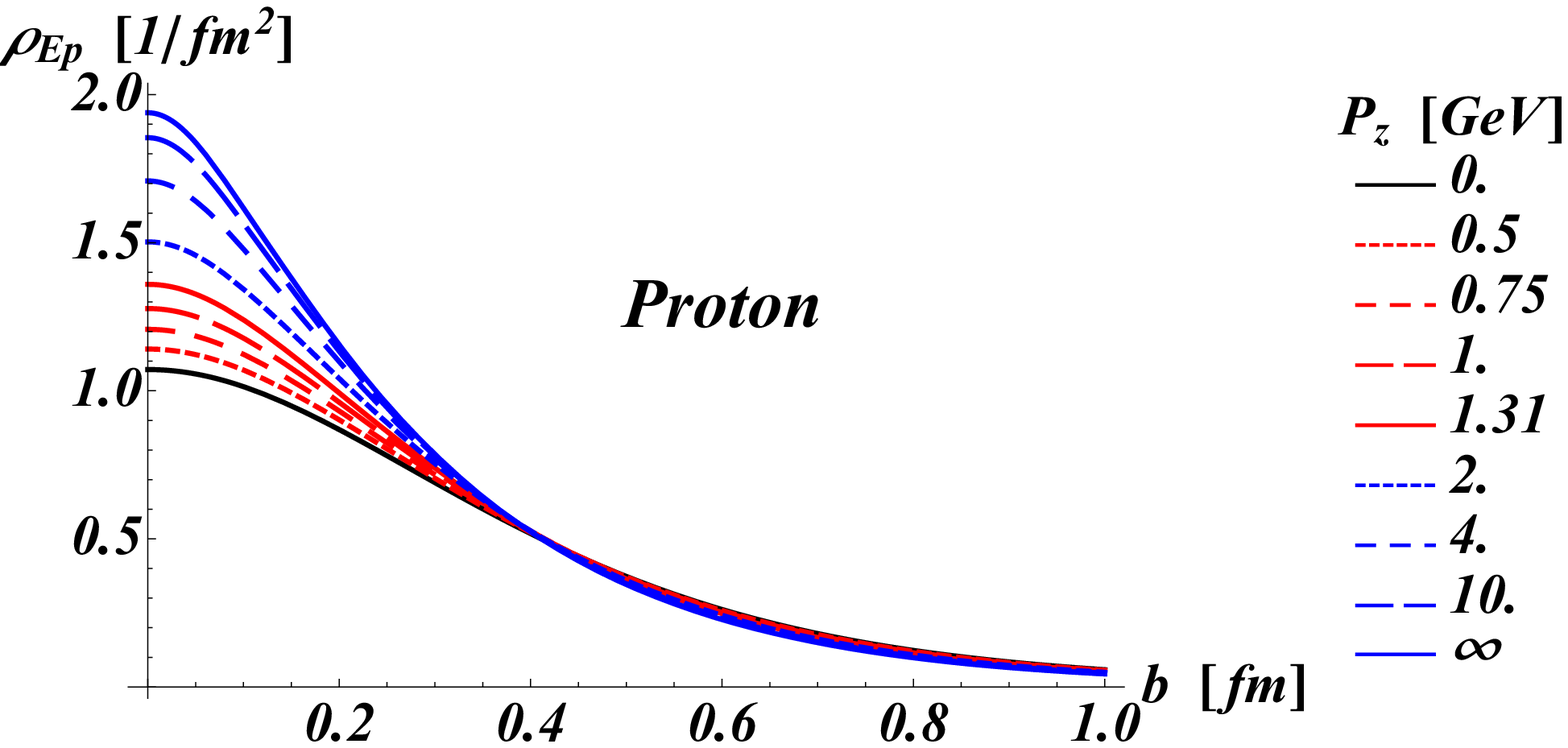}
\caption{(color online) Unpolarized proton 2D charge quasi-density as a function of $P_z$ (lower panel), decomposed into convection and magnetization contributions (upper panels). In the Breit or rest frame $P_z=0$, the charge distribution is purely convective. As $P_z$ increases, a large contribution induced by the rest-frame magnetization progressively concentrates the charge distribution towards the center. Based on the parametrization from~\cite{Bradford:2006yz}.}\label{fig:rhop}
\end{figure}

\begin{figure}[t]
\includegraphics[width=0.51\hsize]{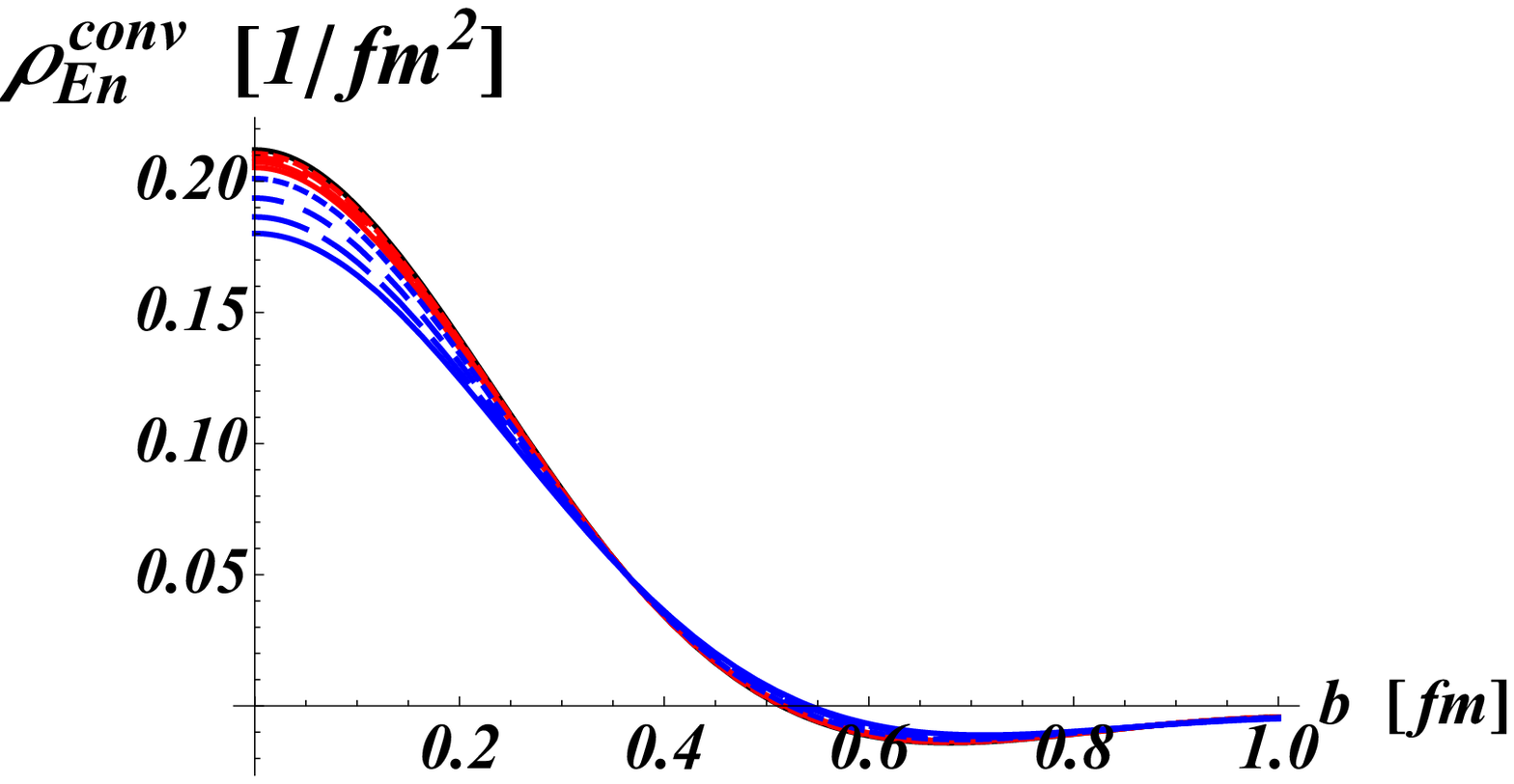}\hspace{-.4cm}
\includegraphics[width=0.51\hsize]{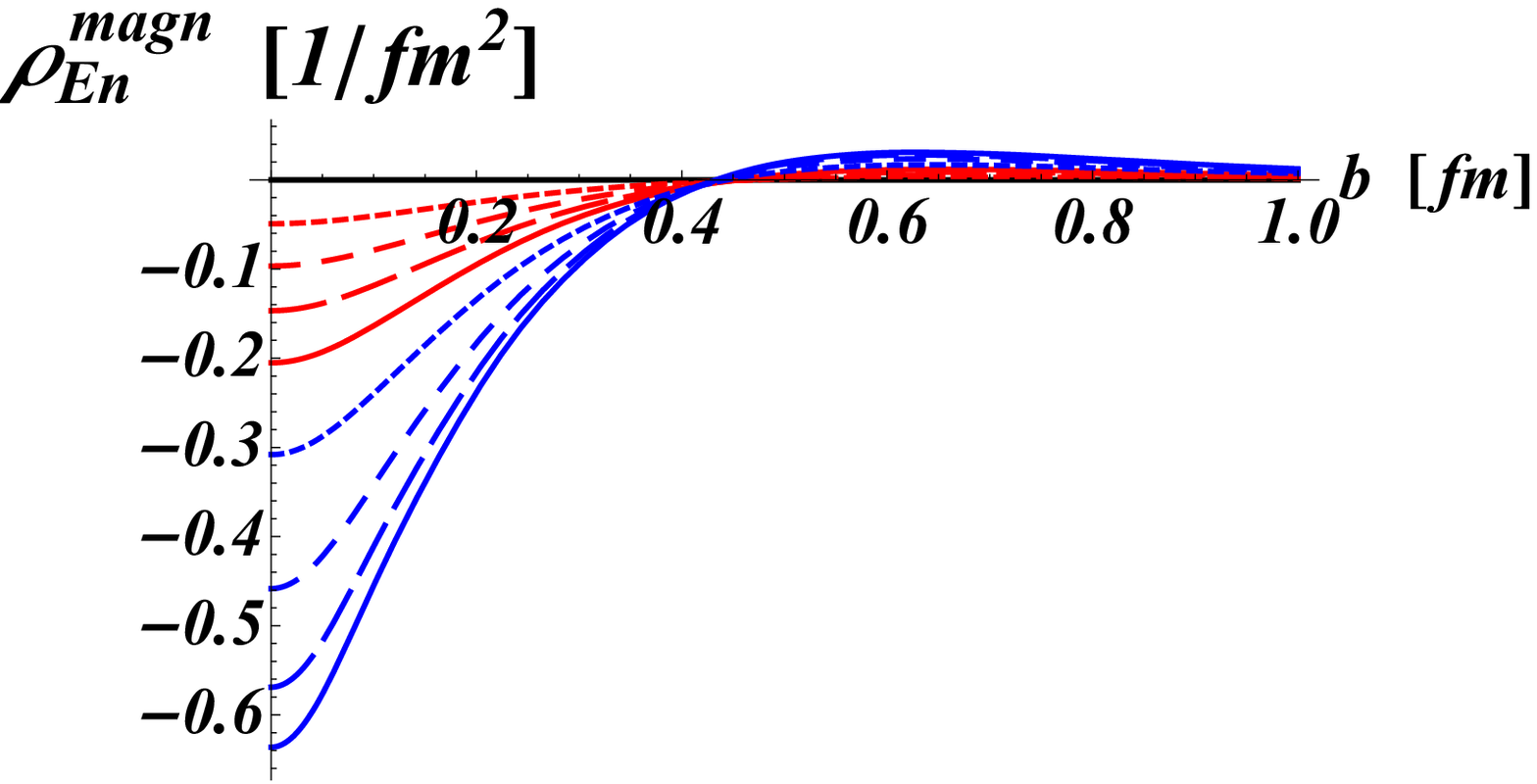}\vspace{.3cm}
\includegraphics[width=0.85\hsize]{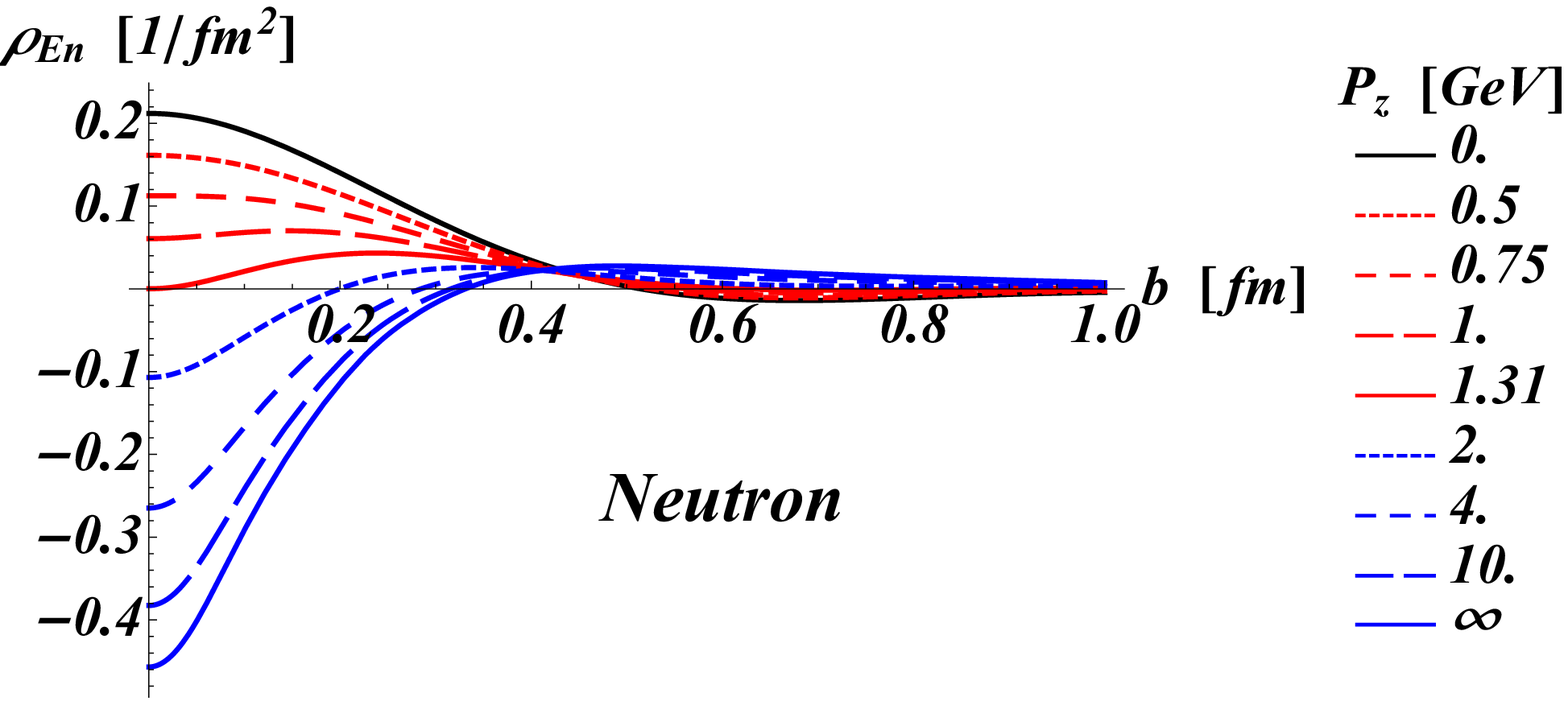}
\caption{(color online) Unpolarized neutron 2D charge quasi-density as a function of $P_z$ (lower panel), decomposed into convection and magnetization contributions (upper panels). In the Breit or rest frame $P_z=0$, the charge distribution is purely convective. As $P_z$ increases, a large contribution induced by the rest-frame magnetization progressively pushes the positive charges away from the center. Based on the parametrization from~\cite{Bradford:2006yz}.}\label{fig:rhon}
\end{figure}

In Figs.~\ref{fig:rhop} and~\ref{fig:rhon} we show how the unpolarized 2D charge quasi-densities of the nucleons evolve with the target momentum $P_z$, using the phenomenological parametrization from~\cite{Bradford:2006yz}. A decomposition into convection and magnetization contributions reveals that the $P_z$-dependence essentially arises from the latter. The mild changes in the convection contribution are entirely due to Wigner rotation effects. The same effects explain why a magnetization contribution, usually associated with transverse polarization, appears in the unpolarized charge distribution. 

In the proton case, the rest-frame magnetization is large and positive. The contribution it induces simply adds up to the convection contribution and increases the charge distribution at the center by almost a factor two.

The situation is more dramatic for the neutron, where the rest-frame magnetization is large and negative. The contribution it induces competes with the convection contribution and gradually changes the sign at the center of the charge distribution. Based on the phenomenological electromagnetic FFs, we find that the center of the charge distribution vanishes when the neutron momentum is around $P_z=1.31$ GeV. 

In summary, we showed that a fully relativistic and model-independent interpretation of the electromagnetic form factors in terms of charge and magnetization distributions can be given within a phase-space approach. Relativistic spatial distributions are quasi-densities and become strict densities only when Galilean symmetry applies. We found that the conventional Sachs distributions in the Breit frame require an unambiguous relativistic kinematical correction to justify their interpretation as rest-frame distributions. We also explained that the distortions appearing in the relativistic distributions for a moving target are entirely due to relativistic kinematical effects associated with spin. In particular, the appearance of a negative region around the center the neutron charge distribution in the infinite-momentum frame is just a manifestation of the contribution induced by the rest-frame magnetization.

The author thanks A. Metz and B. Pasquini for stimulating discussions. This work was supported by the Agence Nationale de la Recherche under the Project ANR-18-ERC1-0002.



\begin{thebibliography}{99}

\bibitem{Perdrisat:2006hj} 
  C.~F.~Perdrisat, V.~Punjabi and M.~Vanderhaeghen,
  Prog.\ Part.\ Nucl.\ Phys.\  {\bf 59}, 694 (2007)
  
\bibitem{Arrington:2006zm}
  J.~Arrington, C.~Roberts and J.~Zanotti,
  J. Phys. G \textbf{34}, S23-S52 (2007).

\bibitem{Punjabi:2015bba} 
  V.~Punjabi, C.~F.~Perdrisat, M.~K.~Jones, E.~J.~Brash and C.~E.~Carlson,
  Eur.\ Phys.\ J.\ A {\bf 51}, 79 (2015).

\bibitem{Pacetti:2015iqa} 
  S.~Pacetti, R.~Baldini Ferroli and E.~Tomasi-Gustafsson,
  Phys.\ Rept.\  {\bf 550-551}, 1 (2015).
  
\bibitem{Ernst:1960zza} 
  F.~J.~Ernst, R.~G.~Sachs and K.~C.~Wali,
  Phys.\ Rev.\  {\bf 119}, 1105 (1960).

\bibitem{Sachs:1962zzc} 
  R.~G.~Sachs,
  Phys.\ Rev.\  {\bf 126}, 2256 (1962).
  
\bibitem{Yennie:1957} 
  D.~R.~Yennie, M.~M.~L\'evy and D.~G.~Ravenhall,
  Rev.\ Mod.\ Phys.  {\bf 29}, 144 (1957).

\bibitem{Breit:1964ga} 
  G.~Breit,
  Proceedings of the XII International Conference on High Energy Physics (ICHEP 1964), pp. 985-987 (1966). 

\bibitem{Kelly:2002if} 
  J.~J.~Kelly,
  Phys.\ Rev.\ C {\bf 66}, 065203 (2002).

\bibitem{Fleming:1974af} 
  G.~N.~Fleming,
  Physical Reality \& Math. Descrip., 357 (1974).

\bibitem{Soper:1976jc} 
  D.~E.~Soper,
  Phys.\ Rev.\ D {\bf 15}, 1141 (1977).

\bibitem{Burkardt:2000za} 
  M.~Burkardt,
  Phys.\ Rev.\ D {\bf 62}, 071503 (2000)
  Erratum: [Phys.\ Rev.\ D {\bf 66}, 119903 (2002)].

\bibitem{Miller:2010nz} 
  G.~A.~Miller,
  Ann.\ Rev.\ Nucl.\ Part.\ Sci.\  {\bf 60}, 1 (2010).
  
\bibitem{Burkardt:2002hr} 
  M.~Burkardt,
  Int.\ J.\ Mod.\ Phys.\ A {\bf 18}, 173 (2003).

\bibitem{Carlson:2007xd}
  C.~E.~Carlson and M.~Vanderhaeghen,
  Phys. Rev. Lett. \textbf{100}, 032004 (2008).

\bibitem{Miller:2007uy} 
  G.~A.~Miller,
  Phys.\ Rev.\ Lett.\  {\bf 99}, 112001 (2007).

\bibitem{Wigner:1932eb} 
  E.~P.~Wigner,
  Phys.\ Rev.\  {\bf 40}, 749 (1932).
  
\bibitem{Hillery:1983ms} 
  M.~Hillery, R.~F.~O'Connell, M.~O.~Scully and E.~P.~Wigner,
  Phys.\ Rept.\  {\bf 106}, 121 (1984).
  
\bibitem{Newton:1949cq}
  T.~Newton and E.~P.~Wigner,
  Rev. Mod. Phys. \textbf{21}, 400-406 (1949).
  
\bibitem{Pavsic:2017orp}
  M.~Pavsic,
  Adv. Appl. Clifford Algebras \textbf{28}, no.5, 89 (2018).
  
\bibitem{Lorce:2017wkb} 
  C.~Lorc\'e, L.~Mantovani and B.~Pasquini,
  Phys.\ Lett.\ B {\bf 776}, 38 (2018).

\bibitem{Lorce:2018egm} 
  C.~Lorc\'e, H.~Moutarde and A.~P.~Trawi\'nski,
  Eur.\ Phys.\ J.\ C {\bf 79}, no. 1, 89 (2019).

\bibitem{Hand:1963zz} 
  L.~N.~Hand, D.~G.~Miller and R.~Wilson,
  Rev.\ Mod.\ Phys.\  {\bf 35}, 335 (1963).

\bibitem{Friar:1975pp}
  J.~L.~Friar and J.~W.~Negele,
  ``Theoretical and Experimental Determination of Nuclear Charge Distributions,''
  In: Baranger M., Vogt E. (eds), Advances in Nuclear Physics, pp. 219-376, Springer, Boston, MA (1975).

\bibitem{Bradford:2006yz}
  R.~Bradford, A.~Bodek, H.~S.~Budd and J.~Arrington,
  Nucl. Phys. B Proc. Suppl. \textbf{159}, 127-132 (2006).
   
\bibitem{Jacob:1959at}
  M.~Jacob and G.~C.~Wick,
  Annals Phys. \textbf{7}, 404-428 (1959).

\bibitem{Durand:1962zza}
  L.~Durand, P.~C.~DeCelles and R.~B.~Marr,
  Phys. Rev. \textbf{126}, 1882-1898 (1962).

\bibitem{Moller:1949}
  C.~M\o ller,
  Theor.\ Phys.\ vol. 3. Commun.\ Dublin\ Inst.\ Advanced\ Studies\ A {\bf 5}, 3 (1949).

\bibitem{Lorce:2018zpf}
  C.~Lorc\'e,
  Eur. Phys. J. C \textbf{78}, no.9, 785 (2018).

\bibitem{Lorce:2011zta}
  C.~Lorc\'e and B.~Pasquini,
  Phys. Rev. D \textbf{84}, 034039 (2011).

\bibitem{Lorce:2017isp} 
  C.~Lorc\'e,
  Phys.\ Rev.\ D {\bf 97}, no. 1, 016005 (2018).

\bibitem{Rinehimer:2009yv} 
  J.~A.~Rinehimer and G.~A.~Miller,
  Phys.\ Rev.\ C {\bf 80}, 015201 (2009).
  
\bibitem{Carlson:2008zc}
  C.~E.~Carlson and M.~Vanderhaeghen,
  Eur. Phys. J. A \textbf{41}, 1-5 (2009).

\bibitem{Alexandrou:2009hs}
  C.~Alexandrou, T.~Korzec, G.~Koutsou, C.~Lorc\'e, J.~W.~Negele, V.~Pascalutsa, A.~Tsapalis and M.~Vanderhaeghen,
  Nucl. Phys. A \textbf{825}, 115-144 (2009).

\bibitem{Lorce:2009bs}
  C.~Lorc\'e,
  Phys. Rev. D \textbf{79}, 113011 (2009).



\end{thebibliography}
\end{document}